\documentclass[aps,twocolumn,showpacs,superscriptaddress]{revtex4}
\usepackage{latexsym}

\usepackage{amsbsy}
\usepackage{bm}
\usepackage{graphicx}

\newcommand{\nc}{\newcommand}
\nc{\da}{\dagger}
\nc{\noi}{\noindent}     
\nc{\eq}[1]{\mbox{Eq.~(\ref{#1})}}
\nc{\ba}{\begin{array}}
\nc{\ea}{\end{array}}
\nc{\bea}{\begin{eqnarray}}
\nc{\eea}{\end{eqnarray}}
\nc{\fig}[1]{\mbox{Fig.~\ref{#1}}}

\nc{\bc}{\begin{center}}
\nc{\ec}{\end{center}}

\begin{document}
\title{Interference of Light in Michelson--Morley 
Interferometer: A~Quantum~Optical~Approach}

\author{{\O}. Langangen}
\email{oysteol@astro.uio.no}
\affiliation{Institute of Theoretical Astrophysics, University of Oslo, 
P.O. Box 1029 Blindern, N-0315 Oslo, Norway}
\affiliation{Centre for Ecological and Evolutionary Synthesis (CEES),
University of Oslo, Department of Biology, P.O. Box 1066 Blindern,
N-0316 Oslo, Norway}
\author{B.-S. Skagerstam}
\email{bo-sture.skagerstam@ntnu.no}
\affiliation{Department of Physics, The Norwegian University of Science and 
Technology, N-7491 Trondheim, Norway}
\author{A. Vaskinn}
\email{asle.vaskinn@ntnu.no}
\affiliation{Department of Physics, The Norwegian University of Science and 
Technology, N-7491 Trondheim, Norway}


\begin{abstract}

  We investigate how the temporal coherence interference properties of light in a
   Michelson-Morley interferometer (MMI), using only a single-photon detector, can be understood in a quantum-optics framework in a straightforward and pedagogical manner. For this purpose we make use of elementary quantum field theory  and Glaubers 
  theory for photon detection in order
  to calculate the expected interference pattern 
  in the MMI. If a thermal reference source is used in the MMI local oscillator port
  in combination with a thermal source in the signal port, the  
  interference pattern revealed by such an intensity measurement shows a distinctive dependence on the differences in the
  temperature of the two sources. The MMI can therefore be used in order to perform
  temperature measurements. A related method was actually used to carry out high precision
  measurements of  the cosmic micro-wave background radiation on board of the COBE  satellite.
  The theoretical framework allows us to consider any initial  {\it quantum }
 state. The interference of single photons as a tool to determine the peak angular-frequency of a one-photon pulse interfering with a  single-photon reference pulse is, e.g., considered.
A similar consideration for laser pulses, in terms of coherent states, leads to a  different response in the detector. The MMI experimental setup is therefore in a sense an example of an optical device where one can exhibit the difference between classical and quantum-mechanical light using only  intensity measurements.
\end{abstract}
\maketitle
\vspace{1cm}
\section{Introduction}
\label{sec:intro}
In 2006 G. Smooth and J. Mather shared the Nobel Prize in Physics
"for their discovery of the black-body form and anisotropy of the 
cosmic micro-wave background radiation (CMB)" \cite{KVA_2006}. These exciting
discoveries were a breakthrough in modern cosmology by the  CMB anisotropy
and the strong validation of the black body spectrum as predicted by the Big Bang theory.
The discovery of the black body form of the CMB spectrum and the 
high precision measurement of the CMB temperature (see e.g. Ref.\cite{Mather_2002} )
relied heavily on the so called Far Infrared Absolute Spectrophotometer (FIRAS)
\cite{Mather_1993} on board the Cosmic Background 
Explorer (COBE)
\cite{Mather_1982, Galileo_2009}.  
In short the FIRAS is a Michelson--Morley interferometer 
enabling a comparison of the interference patterns between an observed 
source and a reference source on-board the COBE satellite.

In this paper we will make use of Glaubers theory for photon detection 
\cite{Glauber_1963,Glauber_1965} (for a guide to the early literature see e.g. Refs.\cite{Klauder_1985} and for  text-book accounts see e.g. Refs.\cite{Klauder_1968, Mandel_1995, Scully_1997}) together with elementary quantum  mechanics to show how the principles of the FIRAS can be
understood in a quantum-optics framework. 

In Section we \ref{sec:Inter} recapitulate the principles of Glaubers photon detection theory
and the transformation properties of a quantum field in a beam-splitter (see e.g. Refs.\cite{Klauder_1986,Mandel_1989,Teich_1989}). The Glauber theory of optical coherence is by now well established and plays a central role in fundamental studies of quantum interference effects of photon quantum states (see e.g. Refs.\cite{Mandel_1999,Zeilinger_1999,Zeilinger_2005}).  In Section \ref{sec:FockCoh} we consider temporal interference effects in
the Michelson--Morley interferometer for pure quantum states like single-photon states as well as classical states corresponding to coherent states (see e.g.  Refs.\cite{Klauder_1985}). 
In Section \ref{sec:Thermal} we explain the principles of interference of thermal light in
the Michelson--Morley interferometer by using only
vacuum as the reference source and reproduce known expressions   and, in Section \ref{sec:cobe},  we consider the full system with an observed thermal source combined with
a thermal reference source.  With the results obtained we then explain the basic principle of FIRAS 
and how it was used as a high precision thermometer. In Section \ref{sec:conclusions} we, finally, give  some concluding remarks. \\
\indent Our presentation extends a recent presentation by Donges \cite{Donges_1998}  and illustrates, e.g.,  that a quantum-mechanical treatment directly leads to the concept of a thermal coherence length without explicitly making use of classical results like the Wiener-Kintchine theorem as in Ref.\cite{Donges_1998}. 
\section{The Michelson-Morley interferometer}
\label{sec:Inter}
\begin{figure}[htb]
\vspace{0.5cm}
\includegraphics[width=0.5\textwidth]{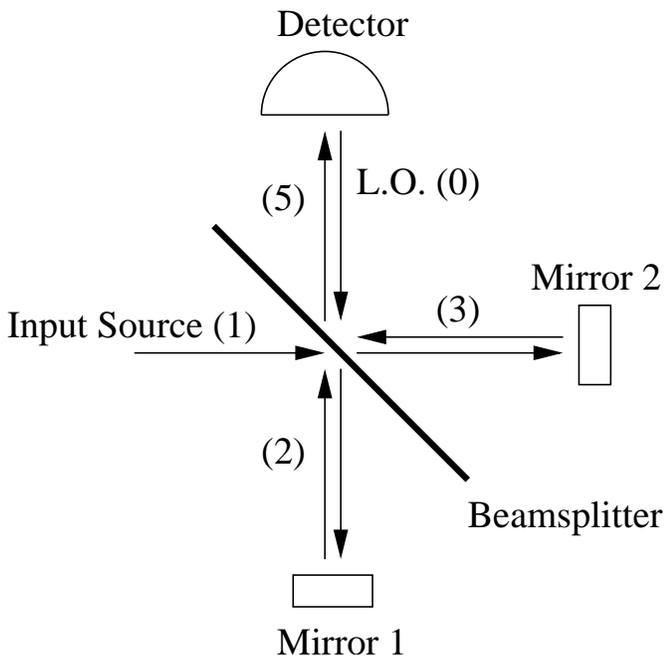}
\caption{Schematic drawing of the MMI setup considered in this paper.
The thick and diagonal  line is representing a beam-splitter with a
transmission coefficient $T$. The numbers in the parenthesis are referring
to the indices use to describe the different light beams  in the calculations.
The signal port (1) and the local oscillator port LO (0) are prepared with various quantum states, like Fock states, coherent states or states with random phases like thermal states. In order to describe  photon absorption in the detector we make use of Glaubers photon detection theory.
}
\label{setup}
\vspace{0.5cm}
\end{figure}

We consider the Michelson-Morley interferometer (MMI) as illustrated in Fig.\ref{setup}, where the so called {\sl temporal} coherence properties of the radiation field is probed (for an early account see e.g. Ref.\cite{Mandel_1965}).
In order to understand the appearance of  interference effects in the MMI, we first discuss the  separate 
parts  of the MMI before we consider the full setup with
the presence of a reference beam.
\subsection{Glaubers theory of photon detection}
Let us first outline a simple, but not unrealistic, model of a photon detector situated at the space-time point $(\vec{x},t)$. 
In this simplified model of a photon detection process \cite{Glauber_1963,Glauber_1965},
the detection of the photon is described by an annihilation of a photon at the detector which modifies the initial state as follows:
\begin{equation}
 {\mid}i{\rangle} \rightarrow  \vec{E}^{(+)}(\vec{x},t){\mid}i{\rangle}\,.
\end{equation} 
The observable electric field operator  $\vec{E}(\vec{x},t)=  \sum_{m}\vec{E}_{m}(\vec{x},t) $ is described in terms of a suitable normal mode expansion,  indexed by mode the number $m$, as a superposition of positive and negative angular-frequency  contributions:
\begin{equation}
\vec{E}_{m}(\vec{x},t)=\vec{E}_{m}^{(+)}(\vec{x},t)+
\vec{E}_{m}^{(-)}(\vec{x},t)\,,
\end{equation}
where $\vec{E}_{m}^{(+)}(\vec{x},t)$ ($\vec{E}_{m}^{(-)}(\vec{x},t)$) contains an annihilation (creation) operator for a photon with mode number $m$.
According to the basic Born rule  in quantum mechanics, the probability to detect the system in a final state $ {\mid}f{\rangle}$, after the one-photon absorption process, is then proportional to $|\langle f {\mid}\vec{E}^{(+)}(\vec{x},t){\mid}i{\rangle}|^2$. Since the exact details of the final states are, in general, unknown we  sum over all possible final states $ {\mid}f{\rangle}$.  In general, we also have to consider not only a pure initial quantum state but also   a quantum state as described by a density matrix. This leads to a description
of the observed intensity $I$ which we write in form
\begin{equation}
I =\mbox{Tr}\Big[\rho\vec{E}^{(-)}(\vec{x},t)
\vec{E}^{(+)}(\vec{x},t)\Big]\,,
\label{photon_detect}
\end{equation}
where  $\rho$ is the density matrix describing the initial state, and where use have be made of the completeness of all possible final states, i.e.  $\sum_f |f \rangle \langle f| = {\bf 1} $. 
 Since we will not be interested in the absolute normalization of the observed intensity $I$, we can neglect normalization constants that may enter into  $I$. It is a remarkable fact that an analysis of single-photon interference in  a Young interferometer using such a quantum-mechanical description of the photon detection process is fairly recent in the history physics
 \cite{Walls_1977} as well as a proper experimental investigation of single-photon interference
 \cite{Grangier_1986,Aspect_1987}.

For reasons of clarity, we will now consider a normal mode expansion of the electro-magnetic field observable $\vec{E}(\vec{x},t)$ in terms of plane waves, i.e. in terms of a mode sum over   wave-vectors $\vec{k}$ and
polarization-vector   labels $\lambda$, i.e. we have for its positive angular-frequency part that
\begin{equation}
\label{evec0}
\vec{E}^{(+)}(\vec{x},t)=i\sum_{\vec{k}\lambda}\sqrt{\frac{\hbar\omega_k}{2V\epsilon_0}}
\vec{\epsilon}_{\vec{k}\lambda}a_{\vec{k}\lambda}e^{i\vec{k}\cdot\vec{x} -i\omega_{k} t}\,.
\end{equation}
Here $\omega_k=c|\vec{k}|$ and  $t$ is a suitable retarded time parameter which will be given in terms of the time-of-flight, given the optical paths in the MMI setup. If the direction of the light beams considered  are well-defined, the dependence of  the detector position $\vec{x}$  can be neglected in the expression for $I$. In general spatial modulations of the measured intensity  are expected \cite{Herzog_94,Shih_2004}. A theoretical analysis of such effects along lines as discussed in the literature (see e.g. Refs. \cite{Wang_91,Milonni_96}) will, however,  not enlighten the issues we are addressing in the present paper. 
$V$ is a quantization volume that will be allowed to be arbitrarily large at an appropriate
late stage of our calculations.  $a_{\vec{k}\lambda}$ is 
the annihilation operator describing the detected light  and $\vec{\epsilon}_{\vec{k}\lambda}$ denotes
the unit polarization vector of the normal mode considered.

Since, in the end,  the dependence of normalization constants will be irrelevant, and since we will only consider polarization-independent optical devices, we make use of a scalar notation. We therefore suppress the wave-vector and polarization labels and with $\omega \equiv \omega_k$ we write $a(\omega) \equiv a_{\vec{k}\lambda}$, such that $[a(\omega) ,a^{\dagger}(\omega) ]=\delta_{\omega \omega'}$ in terms of a discrete Kronecker delta $\delta_{\omega \omega'}$. We also make use of the following convenient notation \cite{Mandel_1995} for the positive angular-frequency part of the electric field at the position of the detector at time $t$ 
\begin{equation}
\label{evec}
E^{(+)}(t) =i\sum_{\omega}\sqrt{\delta \omega }\omega^{1/2}
a(\omega)e^{i\phi (t)}\, .
\end{equation}
Here $\phi (t) \equiv - \omega (t - \tau_s)$ now denotes an optical phase which explicitly takes the source-detection retardation time  into account by the time-delay $\tau_s$, which will be evaluated for the MMI-setup below. If the detector time $t$ enters explicitly into the detection intensity Eq.(\ref{photon_detect}), we will perform a time-average which corresponds to a finite detector time-resolution window. The corresponding time-average of the observed intensity $I \equiv I(t)$ will be denoted by $\langle I \rangle$, i.e.
\begin{equation}
\label{aveint}
\langle I \rangle = \frac{1}{T_{int}}\int_{-T_{int}/2}^{T_{int}/2}dtI(t)\, ,
\end{equation}
where the time $T_{int}$ of integration, as e.g. the time during which an actual measurement proceeds, is chosen to be sufficiently large in comparison with any reciprocal bandwidth of the initial quantum states $ {\mid}i{\rangle}$ considered. The time-averaged observed intensity $\langle I \rangle$ will in general, as we will see explicitly below,  be a function of a time-delay $\tau$ depending on the actual experimental set-up.

  For a finite quantization volume $V$, $\omega$ can be regarded to be discrete and, in the infinite volume limit, we make use of the rule that $\sum_{\omega}\delta \omega \rightarrow \int_{0}^{\infty}d\omega\omega ^{d-1}$, where $d$ is the number of space-dimensions. We will neglect the dependence of transverse dimensions of very collimated normal modes and assume that $d=1$. When appropriate we will, however,  also consider $d=3$ in order to compare with related results in the literature \cite{Donges_1998,Mandel_1965}. Our main results will, however,  not be very sensitive to the choice of $d$.

%
%

Using the same notation as above, a one-photon quantum state ${\mid}f{\rangle}$, with an angular-frequency distribution given by $f\equiv f(\omega)$, is then given by
\begin{equation}
\label{one_photon}
{\mid}f{\rangle} = \sum_{\omega}\sqrt{\delta \omega } 
f(\omega){\mid}1_\omega{\rangle} \equiv (f,a^\dagger){\mid}0{\rangle} \,,
\end{equation}
where ${\mid}1_{\omega}{\rangle} =a^{\dagger}(\omega) {\mid}0{\rangle}$ denotes a one-photon state with angular-frequency $\omega$, normalized according to $\langle 1_{\omega} {\mid}1_{\omega '}{\rangle} =\delta_{\omega \omega'}$, and  where ${\mid}0{\rangle}$ denotes the vacuum state. We also make use of the notation $(f,a^\dagger) \equiv \sum_{\omega}\sqrt{\delta \omega } 
f(\omega)a^{\dagger}(\omega)$. The state ${\mid}f{\rangle}$ above is an eigenstate of the number operator $\hat{N} =\sum_{\omega}a^{\dagger}(\omega)a(\omega)$, i.e. a Fock state, with, of course, an eigenvalue corresponding to one particle present. Normalization of the state ${\mid}f{\rangle}$ for $d=1$
therefore corresponds to 
\begin{equation}
\label{one_normalization}
\langle f {\mid}f{\rangle}=1=\sum_{\omega}\delta \omega |f(\omega)|^2 = \int_{0}^{\infty}d\omega  |f(\omega)|^2 \, ,
\end{equation}
in the large-volume $V$ limit.
In order to make our presentation quantitative  we will, for reasons of simplicity, consider  real-valued one-photon angular-frequency distributions $f(\omega)$ such that
\begin{equation}
\label{f_states}
f(\omega)  = \frac{1}{N}\exp(-(\omega -{\bar \omega} )^2/2\sigma^2)\, ,
\end{equation}
with a mean angular-frequency ${\bar \omega}$ and width $\sigma$ and where the normalization constant $N$ is given by 
\begin{equation}
\label{norm_state}
|N|^2 = \frac{\sigma\sqrt{\pi}}{2}\left( 1+ \frac{2}{\sqrt{\pi}}\int_{0}^{{\bar \omega}/\sigma}dx e^{-x^2} \right)
\end{equation}
in terms of an error function.
This choice of frequency distribution makes it possible to actually carry out all relevant expressions analytically. In obtaining the properly normalized expression  Eq.(\ref{f_states}) we keep $\omega \ge 0$. It may, however,  sometimes be possible to extend the range of angular frequencies to arbitrarily negative values in Eq.(\ref{one_normalization}), so that $|N|^2= \sigma \pi$, with an exponential small error, which makes some of the expressions as given below more tractable and transparent. With our choice of beam parameters below it turns out that such an  approximation plays only a minor role with regard to actual numerical evaluations. 

Conventional coherent states ${\mid}f{\rangle}_c$,   as expressed in terms of  the one-photon distribution $f$, can then be obtained using a multi-mode displacement operator (see e.g. Refs.\cite{Klauder_1985}), i.e. 
\begin{equation}
\label{coherent_state}
{\mid}f{\rangle}_c = e^{(f,a^\dagger)-(f^*,a)} {\mid}0{\rangle} = e^{-\langle f{\mid}f\rangle/2} e^{(f,a^\dagger)}{\mid}0{\rangle} \, .
\end{equation}
such that 
\begin{equation}
a_{\omega}{\mid}f{\rangle}_c =  (\delta\omega)^{1/2}f(\omega) {\mid}f{\rangle}_c\, ,
\end{equation}
and hence
\begin{equation}
_{c}\langle f{\mid} \hat{N}{\mid}f{\rangle}_c = \sum_{\omega}\delta\omega |f(\omega )|^2 .
\end{equation}
\subsection{Transformation in the beam-splitter}
Next, we consider a beam-splitter with frequency independent transmittance $T$
and reflectance $R$. If we assume a prefect beam-splitter, where
all light is either reflected or transmitted, we have $R+T=1$. 
The input annihilation operators $a_0(\omega)$ and $a_1(\omega)$ of the beam-splitter will then transform according to (see e.g. Refs.\cite{Klauder_1986,Mandel_1989,Teich_1989})
\begin{eqnarray}
a_2(\omega) &=&\sqrt{T}a_0(\omega) +i\sqrt{1-T}a_1(\omega) \,,\nonumber\\
a_3(\omega) &=&\sqrt{T}a_1(\omega) +i\sqrt{1-T}a_0(\omega) \,,
\label{bs}
\end{eqnarray}
where $a_0(\omega),a_1(\omega)$ are the 
L.O. and  signal port mode annihilation operators, and $a_2(\omega),a_3(\omega)$ the output annihilation mode operators corresponding to the transmitted and reflected  modes, respectively.
The $\pi/2$--phase-shift between the transmitted and reflected part,
described by the complex numbers in Eq.(\ref{bs}) will play and important role below as is also the case in the famous Hong-Ou-Mandel two-photon correlation experiment \cite{Hong_Mandel} and related investigations (see e.g. Ref.
\cite{Shih_1988,Chiao_93,Strekalov_1998}).

%
%
\begin{figure}[htp]
\vspace{1.0cm}
\leftline{ \includegraphics[width=0.9\columnwidth]{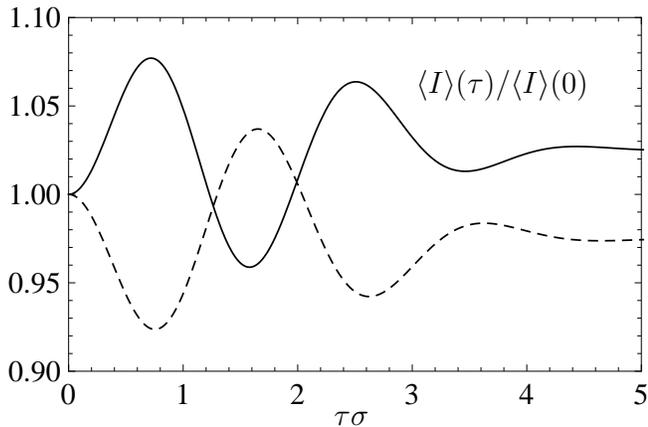}}
\caption{The normalized single-photon intensity $\langle I \rangle (\tau)/\langle I \rangle (0)$
as a function of the dimensionless time-delay $\tau \sigma$ for the case of
one-photon states in the signal and  the LO ports with the same spectral width $\sigma$ but with different mean frequencies.  Two different examples are plotted with 
${\bar \omega}_s= 3\sigma$:  ${\bar \omega}_{lo}= 3.15 \sigma$ with (solid line) or 
${\bar \omega}_{lo}= 2.85 \sigma$ (dashed line). The asymptotic values of $\langle I \rangle (\tau)/\langle I \rangle (0)$ can be obtained from the expression Eq.(\ref{fockave_approx}) in the main text, i.e. $(1 + {\bar \omega}_{lo}/{\bar \omega}_{s})/2$.}
\label{Fock_Fig}
\vspace{1.0cm}
\end{figure}
%

A light beam arriving at the beam-splitter after being reflected in 
the mirrors $1$ and $2$  will, in general, be phase shifted, i.e. expressed in terms of mode operators this process corresponds to the time-of-flight replacement
\begin{eqnarray}
a_2(\omega)&\rightarrow&a_2(\omega)e^{i\phi_2(t)}\,,\nonumber\\
a_3(\omega)&\rightarrow&a_3(\omega)e^{i\phi_3(t)}\,,
\end{eqnarray}
due to difference in optical path lengths with $\phi_2=-\omega(t-\tau_2),\,
\phi_3=-\omega(t-\tau_3)\,$ in terms of  time-delays $\tau_2$ and $\tau_3$. The reflections at the mirrors in the MMI setup will also
introduce a phase-shift of $\pi/2$, but this is equal for the two light beams and we can therefore be neglected all together.
A light beam passing through the beam-splitter after reflection at the mirrors will, again, transform according to Eq.(\ref{bs})
and we therefore, finally, obtain an expression for the mode operator describing  incident light on the photon detector, i.e.
\begin{eqnarray}
a_5(\omega)= \sqrt{T} a_2(\omega)e^{i\phi_2(t)}+i\sqrt{1-T}a_3(\omega)e^{i\phi_3(t)}
\nonumber\\
= a_0(\omega)\left( T e^{i\phi_2(t)}-(1-T)e^{i\phi_3(t)}\right)+\nonumber~~~~~~\\
a_1(\omega)\left(i\sqrt{T(1-T)}e^{i\phi_2(t)}+
i\sqrt{T(1-T)}e^{i\phi_3(t)}\right)\,.
\end{eqnarray}
The positive angular-frequency part of the electro-magnetic field operator, $E^{(+)}(t)$, at the position of the detector  as given in Eq.(\ref{evec}),  will now be expressed in terms of $a_5 (\omega)$. For reasons of clarity we will consider a $50/50$ beam splitter, i.e. we make the choice $T=1/2$, and therefore the corresponding electro-magnetic field operator  to be used in Glaubers theory of photon detection, is given by
\begin{eqnarray}
\label{edetect}
E^{(+)}(t) = i\sum_{\omega}\frac{1}{2}\sqrt{\delta \omega }\omega^{1/2}
\left( a_0(\omega)(e^{i\phi_2(t)}-e^{i\phi_3(t)}) + \right. \nonumber \\
 \left. ia_1(\omega)(e^{i\phi_2(t)}+e^{i\phi_3(t)})\right) \, .~~~~~~~~~~~~~~~~
\end{eqnarray}
\section{Interference of Fock States and Coherent States}
\label{sec:FockCoh}
Let us now specifically consider the following initial Fock state
\begin{eqnarray}
\label{eq:fockintial}
\mid i \rangle = {\mid} f_s \rangle\otimes{\mid}f_{lo} \rangle  \,\, , 
\end{eqnarray}
for the signal and the local port with one-photon angular-frequency distributions  according to Eq.(\ref{f_states})  with the same spectral widths $\sigma $ but with $\bar{ \omega} \equiv \bar{ \omega }_s =3 \sigma$ for the signal port  and $\bar{ \omega} \equiv  \bar{ \omega }_{lo} = 3.15 \sigma$ or $\bar{ \omega }_{lo} =2.85 \sigma$ for the local oscillator port.
We then see that
\begin{eqnarray}
\label{e_on_i}
E^{(+)}(t){\mid}i \rangle =~~~~~~~~~~~~~~~~~~~~~~~~~~~~ \nonumber \\
i\sum_{\omega}\frac{1}{2}\delta \omega \omega^{1/2}
\left( i f_s(\omega)e^{i\phi_2(t)}(1+e^{i\omega\tau}){\mid}0\rangle\otimes{\mid}f_{lo} \rangle \right.~~~~~~\nonumber \\
 \left.   +  
  f_{lo}(\omega)e^{i\phi_2(t)}(1-e^{i\omega\tau}) ){\mid}f_s \rangle\otimes{\mid}0 \rangle\right) \, ,~~~~~
\end{eqnarray}
where  we have made use of the fact that $\phi_3(t)-\phi_2(t)= \omega(\tau_3-\tau_2) \equiv \omega \tau$, which now defines the optical time-delay $\tau$. 
The probability for the detection of one photon will  according to Glaubers theory of photon detection, as we have mentioned above, be proportional to the absolute modulus square of the probability amplitude  $E^{(+)}(t){\mid}i \rangle$, i.e. to $|E^{(+)}(t){\mid}i \rangle|^2$. A time-average over the time $t$ according to Eq.(\ref{aveint}) with $T_{int} \gg 1/\sigma$ leads to a Kronecker delta $\delta_{\omega\omega' }$ and therefore makes any double-sum over frequencies into a single-sum.  In the large-volume limit and for $d=1$, we then obtain 
\begin{eqnarray}
\label{fockave}
\langle I \rangle (\tau) = \frac{1}{2T_{int}}\int_{0}^{\infty}d\omega 2\pi \left( \omega |f_s(\omega)|^2(1 + \cos\omega\tau)~ + \right. \nonumber \\ 
\left. \omega |f_{lo}(\omega)|^2 (1-\cos\omega\tau) \right) \, . ~~~~
\end{eqnarray}
If $\bar{ \omega }_s,\bar{ \omega }_{lo} \gg \sigma$ then, within a good numerical approximation, we can first replace  the linear $\omega$ dependence in Eq.(\ref{fockave}) with  $\bar{ \omega }_s$ and $\bar{ \omega }_{lo}$ in front of the corresponding angular-frequency distributions, and then extend  the integration to include arbitrarily negative angular-frequencies. One then finds that
\begin{eqnarray}
\label{fockave_approx}
\frac{\langle I \rangle (\tau)}{\langle I \rangle (0)}\simeq  \frac{1}{2} \left( 1 + \frac{\bar{\omega }_{lo}}{\bar{\omega }_s}~ +~~~~~~~~~~   \nonumber \right. \\ \left. e^{-(\sigma \tau )^2/4 }\left( \cos\tau\bar{\omega }_s - \frac{\bar{\omega }_{lo}}{\bar{\omega }_s}\cos\tau\bar{\omega }_{lo}\right) \right)  \, , ~~~~
\end{eqnarray}
where the interference effects are exponentially sensitive to the spectral width $\sigma$ of the 
single-photon angular-distributions similar to the spectral width dependence in the famous 
 Hong-Ou-Mandel two-photon experiment \cite{Hong_Mandel}. The asymptotic value of $\langle I \rangle (\tau)/\langle I \rangle (0)$ is given by $(1 + \bar{\omega }_{lo}/\bar{\omega }_{s})/2$. 
 In Fig.\ref{Fock_Fig} we exhibit $\langle I \rangle (\tau)/ \langle I \rangle (0)$ according to Eq.(\ref{fockave}) with the choice as in Eq.(\ref{eq:fockintial}). It is now clear that with a given reference distribution $|f_{lo}(\omega)|^2$ of the local oscillator one can, e.g., infer the  common spectral width $\sigma$ of the single-photon sources  as well as the corresponding angular-frequency $\bar{\omega }_{s}$.  A related experimental situation is discussed in Ref.\cite{Wasilewski_2007} for a general   one-photon  state, i.e. not necessarily a pure quantum state.

\begin{figure}[htb]
\vspace{1.0cm}
\leftline{ \includegraphics[width=0.9\columnwidth]{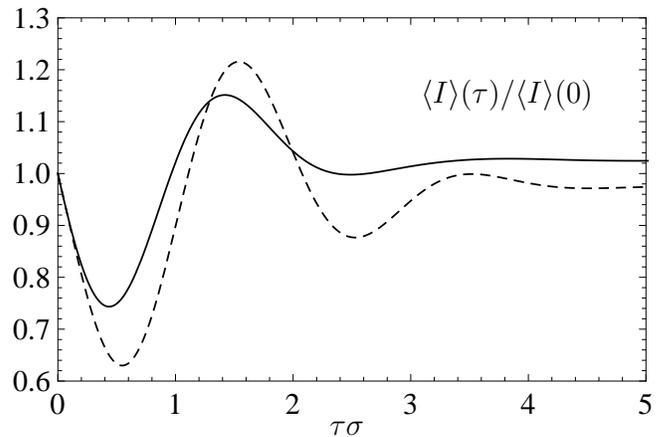}}
\caption{The normalized intensity $\langle I\rangle (\tau)  /\langle I \rangle(0)$
as a function of the time-delay $\tau/\sigma$ for the case of coherent states as generated by the same one-photon states as in Fig.\ref{Fock_Fig}.  Two different examples are plotted with 
${\bar \omega}_s= 3\sigma$:  ${\bar \omega}_{lo}= 3.15 \sigma$ with (solid line) or 
${\bar \omega}_{lo}= 2.85 \sigma$ (dashed line). The asymptotic values of $\langle I\rangle (\tau)  /\langle I \rangle(0)$ are as in Fig.\ref{Fock_Fig}.}
\label{Coh_Fig}
\end{figure}

In the case of coherent states in the signal and local ports, with one-particle state angular-frequency distributions $f_s(\omega)$ and $f_{lo}(\omega)$  as above, Eq.(\ref{e_on_i}) is now modified according to 
\begin{eqnarray}
\label{ce_on_ci}
E^{(+)}(t){\mid}i \rangle = i\sum_{\omega}\frac{1}{2}\delta \omega \omega^{1/2}
\left( i f_s(\omega)e^{i\phi_2(t)}(1+e^{i\omega\tau}) \right.~~~~~~\nonumber \\
 \left.   +  
  f_{lo}(\omega)e^{i\phi_2(t)}(1-e^{i\omega\tau}) \right){\mid}f_s \rangle_c\otimes{\mid}f_{lo} \rangle_c \, ~,~~~~~~~~
\end{eqnarray}
i.e. Eq.(\ref{fockave}) is, for real-valued single-photon distributions $f_s(\omega)$ and $f_{lo}(\omega)$, replaced by
\begin{eqnarray}
\label{coh_ave}
\langle I \rangle (\tau) = \frac{1}{2T_{int}}\int_{0}^{\infty} d\omega \omega 2\pi \left(  |f_s(\omega)|^2(1 + \cos \omega\tau)~ + \right. \nonumber \\ 
\left.  |f_{lo}(\omega)|^2 (1-\cos\omega\tau) - 2f_s(\omega)f_{lo}(\omega)\sin\omega\tau  \right) \, . ~~~~
\end{eqnarray}
In Fig.\ref{Coh_Fig} we exhibit $\langle I \rangle (\tau)/ \langle I \rangle (0)$ with coherent states generated by the  same choice of single-photon states as in Fig.\ref{Fock_Fig}. The interference is, as in Fig.\ref{Fock_Fig}, sensitive to the actual angular-frequency distributions. With a given reference distribution $|f_{lo}(\omega)|^2$ of the local oscillator, one can now infer the common spectral width $\sigma$ of the coherent state sources  as well as the corresponding angular-frequency $\bar{\omega }_{s}$.  By a comparison with Fig.\ref{Fock_Fig}, we  conclude that the MMI setup is sensitive to the actual form of the initial quantum states despite the fact that we are only considering  single-photon detection processes.

\section{Interference of thermal light in the Michelson-Morley 
interferometer}
\label{sec:Thermal}

\subsection{Thermal light} 
For the readers convenience, let us first outline a simple and quantum mechanical description of thermal black body radiation at an absolute temperature $T$. Black body radiation is, basically, light with random phases. For a single mode with  angular-frequency $\omega$ and if
  ${\mid}n_{\omega}{\rangle} =a^{\dagger}(\omega)^{n_{\omega}} {\mid}0{\rangle}/\sqrt{n_{\omega}!}$ denotes a $n_{\omega}$-photon state, the density matrix describing the thermal light is given by
\begin{equation}
\rho(\omega)=\sum_{n_{\omega}=0}^{\infty} p_n(\omega){\mid}n_{\omega}{\rangle}{\langle}n_{\omega}{\mid}\,,
\label{thermal_light}
\end{equation}
in terms of the Bose-Einstein distribution 
\begin{eqnarray}
p_n(\omega)&=&\Big(\frac{\bar{n}(\omega,T)}{1+\bar{n}(\omega, T)}\Big)^n\frac{1}{1+\bar{n}(\omega,T)}
\label{dist_thermal_light}\,,
\end{eqnarray}
with 
\begin{eqnarray}
\bar{n}(\omega , T)&=&\frac{1}{\exp(\hbar\omega/k_{B}T)-1}\,.\nonumber
\end{eqnarray}
The state $\rho(\omega)$ corresponds to an extreme value of 
the von Neumann entropy $S$ in units of $k_BT$, i.e.
\begin{equation}
S=-\mbox{Tr}[\rho(\omega)\ln\rho(\omega)]\,,
\end{equation}
subjected to the constraints (see e.g. Refs.\cite{Mandel_1965,Scully_1997})
\begin{eqnarray}
\mbox{Tr}[\rho(\omega)]&=&1\,,\nonumber\\
\mbox{Tr}[a^{\dagger}(\omega)a(\omega)\rho(\omega)]&=&\bar{n}(\omega ,  T) \,.
\label{norm}
\end{eqnarray}
The random, or chaotic, nature of the quantum state $\rho(\omega)$ corresponds to a phase-independent Glauber-Sudarshan ${\cal P}(\alpha)$-representation (see e.g. Refs.\cite{sudarshan63,glauber_p_63,klauder_et_al_65,klauder2007}) in terms of a coherent state, i.e.
\begin{eqnarray}
\rho(\omega) = \int d^2\alpha {\cal P}(\alpha){\mid}\alpha{\rangle}{\langle}\alpha{\mid}\, ,
\label{P_KGS}
\end{eqnarray}
using a single-mode coherent state ${\mid}\alpha{\rangle} =\exp(\alpha a^{\dagger}(\omega)- \alpha^* a(\omega)){\mid}0{\rangle}$. For thermal light one finds that
\begin{eqnarray}
\label{chaotic}
{\cal P}(\alpha) = \frac{1}{\pi\bar{n}(\omega , T )} \exp \left( -|\alpha|^2/\bar{n}(\omega , T) \right)\, ,
\end{eqnarray}
which obeys the normalization condition
\begin{eqnarray}
\mbox{Tr}[\rho(\omega)]=\int d^2\alpha {\cal P}(\alpha) =1\, ,
\end{eqnarray}
as well as
\begin{eqnarray}
\mbox{Tr}[a^{\dagger}(\omega)a(\omega)\rho(\omega)]=\int d^2\alpha {\cal P}(\alpha) |\alpha|^2 = \bar{n}(\omega , T ) \,.
\label{mean_value}
\end{eqnarray}

A multi-mode system 
at thermal equilibrium is then described in terms of a tensor product $\rho (T) =\bigotimes_\omega \rho(\omega)$, where we have performed the replacement $\alpha \rightarrow \alpha(\omega)$ in Eqs.(\ref{P_KGS}) and (\ref{chaotic}). The Glauber-Sudarshan ${\cal P}(\alpha)$-representation for the state $\rho (T)$ is now, in particular, useful in our considerations since the response in a single-photon detector  can be obtained immediately from the previous results for coherent light using an average procedure.

\subsection{Thermal light in the signal port and 
vacuum in the local oscillator port}
\label{thermal_vac}

\begin{figure}[htb]
\vspace{1.2cm}
\leftline{ \includegraphics[width=0.9\columnwidth]{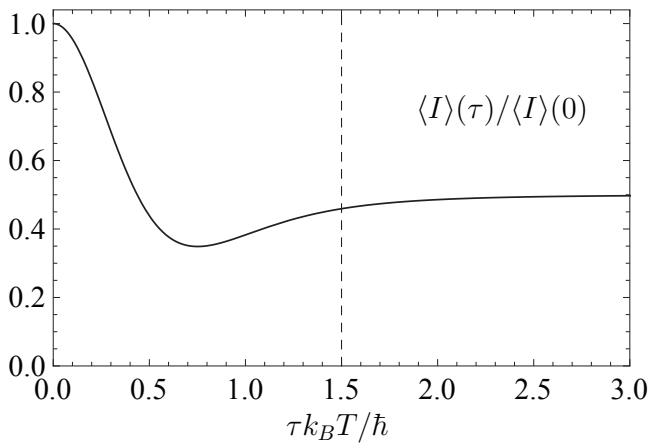}}
\caption{The normalized intensity $\langle I \rangle (\tau)/\langle I \rangle(0)$
as a function of $a \equiv \tau k_BT_0/\hbar$ for the case of
thermal light in the signal port and vacuum in the LO port and for $d=3$. 
We infer a characteristic thermal coherence time $\tau_c $ in terms of $a \simeq 1.5$ (dashed vertical line), i.e. $\tau_c \simeq 1.5\hbar/k_B T$.
}
\label{vac_num}
\vspace{1.0cm}
\end{figure}

We are now in the position to consider a density matrix describing a
system where we have thermal light in the signal port and vacuum in the LO port, i.e. the initial
density matrix $\rho $ of the total system is given by
\begin{eqnarray}
\rho = \rho(T)\otimes ({\mid}0{\rangle}{\langle}0{\mid})_{lo} 
\,.
\label{vacuum}
\end{eqnarray}
By making use of Eq.(\ref{coh_ave}), with $f_s(\omega)\rightarrow \alpha(\omega)$ and $f_{lo}(\omega)\rightarrow 0$, and then performing an average over $\alpha(\omega)$  according to Eq.(\ref{mean_value}), we immediately obtain the result 
\begin{eqnarray}
\label{tempave}
\langle I \rangle (\tau) = \frac{1}{2T_{int}}\int_{0}^{\infty} d\omega 2\pi\omega^d \bar{n}(\omega , T)(1 + \cos\omega\tau ) \, , ~~~~
\end{eqnarray}
in $d$ space dimensions. In passing, we notice that Eq.(\ref{tempave}) is actually valid for {\it any} physical quantum state of the form Eq.(\ref{vacuum}) due to the generality of the Glauber-Sudarshan ${\cal P}(\alpha)$-representation Eq.(\ref{P_KGS}). The interference effects as exhibited by the MMI setup therefore only depends on the, in general,  angular-frequency dependent mean-number  $ \mbox{Tr}[a^{\dagger}(\omega)a(\omega)\rho] =n(\omega)$ and not on other features of the actual quantum state considered. By a straightforward change of integration variables in Eq.(\ref{tempave}),  and with $a\equiv \tau k_BT/\hbar$, we then find that
\begin{equation}
\frac{\langle I \rangle(\tau)}{\langle I \rangle(0)}=\frac{1}{2}\left[ 1 + 
\frac{1}{J(d)}\int_0^\infty dx\Big(\frac{x^d\cos(ax)}
{\exp(x)-1}\Big)\right]\,, 
\label{int_vac}
\end{equation}  
where $J(d)$ can be expressed in terms of gamma and  Riemann's $\zeta$ functions, i.e.
\begin{eqnarray}
J(d) = \Gamma(1+d) \zeta(d+1)= \Gamma(1+d) \sum_{n=1}^{\infty}\frac{1}{n^{d+1}} \, .
\end{eqnarray}
Particular values are $J(1)=\pi^2/6$ and $J(3)=\pi^4/15$. In the case of $d=3$ we, therefore, recover the well-known expression for $\langle I \rangle(\tau)/\langle I \rangle(0)$ \cite{Mandel_1965} as also discussed in Ref.\cite{Donges_1998}.  In, e.g., $d=3$ it is actually possible to carry out the relevant integral in the expression Eq.(\ref{int_vac}) for $\langle I \rangle(\tau)/ \langle I \rangle(0)$ exactly with the result
\begin{equation}
\label{eq_donges}
\frac{\langle I \rangle(\tau)}{\langle I \rangle(0)}=\frac{1}{2}\left[ 1 + 
15\Big( \frac{2+\cosh (2a\pi)}{\sinh(a\pi)^4} - \frac{3}{(a\pi)^4}\Big)\right]\,. 
\end{equation}  
Eq.(\ref{eq_donges}) shows that interference effects have a power-law sensitivity for larger $a$. In Fig.\ref{vac_num} we show $\langle I \rangle(\tau)/\langle I \rangle(0)$
for varying values of $a$ in the case with $d=3$ and  one
infers a characteristic coherence length $l_c$ of thermal light in the MMI of the form 
\begin{equation}
l_c{\simeq}1.5\frac{\hbar c}{k_BT}\,,
\end{equation}
as also discussed in Ref.\cite{Donges_1998}.
%
\section{Thermal light in the signal  and local ports}
\label{sec:cobe}

As we have seen above, with a vacuum in the LO port and with a signal thermal source the single-photon detection process exhibits an  interference pattern.
We now investigate what happens if we have thermal light with a temperature
$T_0$ in the LO port and thermal light with temperature $T_1$ in the signal port.
For this setup the corresponding initial density matrix becomes
\begin{eqnarray}
\rho = \rho(T_1)\otimes \rho(T_0)
\,.
\label{non_vacuum}
\end{eqnarray}
\noindent By making use of this density matrix as well as the
same methods as described in Section \ref{thermal_vac} by performing independent averages over $f_s(\omega)$ and $f_{lo}(\omega)$ in Eq.(\ref{coh_ave}) according to Eq.(\ref{mean_value}), we immediately obtain, in the large-volume limit, the result 
\begin{eqnarray}
\langle I \rangle (\tau) = \frac{1}{2T_{int}}\int_{0}^{\infty} d\omega \omega^d 2\pi \left( \bar{n}(\omega, T_1)(1 + \cos\omega\tau)~ + \right. \nonumber \\ 
\left.  \bar{n}(\omega, T_0)(1-\cos\omega\tau) \right) \, . ~~~~
\label{int_thermal}
\end{eqnarray}
%
%

%
%
%

\begin{figure}[htb]
\vspace{0.5cm}
\leftline{ \includegraphics[width=0.9\columnwidth]{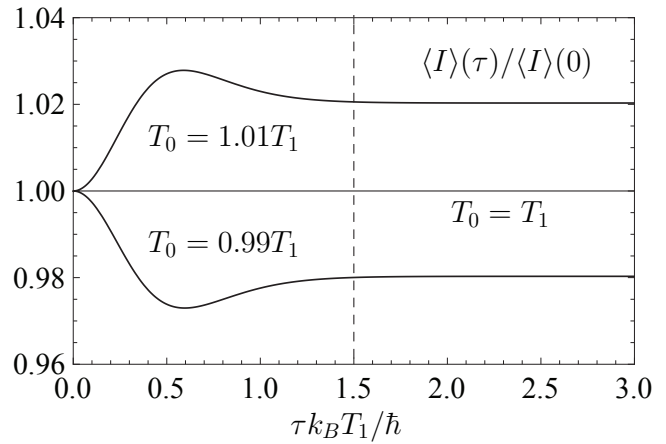}}
\caption{The same as in Fig.\ref{vac_num} but with thermal light
in both ports with
$T_1=1.01T_0$ (dashed-dotted line),
$T_1=0.99T_0$ (dashed line), and  
$T_1=T_0$ (solid line). The vertical dashed line corresponds to
the thermal coherence length $l_c=c\tau_c$ with $\tau_c k_BT_1/\hbar \simeq 1.5$.
}
\label{therm_num}
\end{figure}

\noindent
Here we notice that the last term in Eq.(\ref{coh_ave}), suitably extended to complex-valued $f_s(\omega)$ and $f_{lo}(\omega)$,  averages to zero due to the chaotic nature, i.e. phase-independence,  of thermal light according to Eqs.(\ref{P_KGS}) and (\ref{chaotic}).
The relative intensity with respect to zero time-delay $\tau$ has then the form
\begin{eqnarray}
\frac{\langle I \rangle (\tau)}{\langle I \rangle (0)} =\frac{1}{2}\Big(1 +\big(\frac{T_0}{T_1}\big)^4 \Big)&
+&  \nonumber\\  \frac{15}{2\pi^4}\int_0^\infty dx \frac{x^3}{\exp(x)-1} \Big(\cos(a_1x) &-& 
\big(\frac{T_0}{T_1}\big)^4 
\cos(a_0x) \Big)\,, \nonumber\\
\label{two_therm}
\end{eqnarray}  
where $a_0=\tau k_BT_0/\hbar$  and  $a_1=\tau k_BT_1/\hbar$.
The integrals in Eq.(\ref{two_therm}) can, again, be solved analytically in a 
fashion similar to the integral in Eq.(\ref{int_vac}), i.e.
\begin{eqnarray}
\frac{\langle I \rangle (\tau)}{\langle I \rangle (0)} =\frac{1}{2}\Big(1 &+& \big(\frac{T_0}{T_1}\big)^4 \big( 1-15\frac{2+\cosh (2a_0\pi)}{\sinh (a_0\pi)^4} \big) \Big)
+  \nonumber \\ &&\frac{15}{2}\frac{2 + \cosh (2a_1\pi)}{\sinh (a_1\pi)^4} \, \, . \nonumber\\
\label{two_therm_exact}
\end{eqnarray}  
Due to the presence of hyperbolic functions in Eq.(\ref{two_therm_exact}), we observe that $\langle I \rangle (\tau)/\langle I \rangle (0)$ approaches its asymptotic value $(1+(T_0/T_1)^4)/2$ exponentially fast as a function of the time-delay $\tau$ in contrast to the power-law dependence in Eq.(\ref{eq_donges}).  It is now of interest to study  the behavior of Eq.(\ref{two_therm_exact})
when the  temperature $T_0$ and $T_1$ of the local oscillator and signal respectively are varied. In Fig.\ref{therm_num} we exhibit the interference  when $T_0$ is slightly smaller or larger than 
$T_1$ as a function of the parameter $\tau k_BT_1/\hbar$ for $d=3$. The corresponding interference, of course, disappears
when the two temperatures are equal. The sensitivity  of the 
interference pattern with regard to the difference in temperatures of the source and the reference temperature, i.e. of the local source, constitutes the basic ingredient of the
 FIRAS setup.  We also observe that the coherence length $\l_c$ for
this MMI setup is the roughly same as 
in Section \ref{thermal_vac}.

\section{Conclusions and Final Remarks}
\label{sec:conclusions}
We have seen that the interference of thermal light in the Michelson-Morley 
interferometer  can be described, in a straightforward manner, by making use of Glaubers theory of photon detection 
and elementary quantum theory of the electro-magnetic field. Furthermore, we have 
seen the emergence of a natural coherence length $\l_c \simeq  \hbar c/k_BT$  of thermal light in the MMI. 
It may be surprising that non-monochromatic and chaotic light, with random phases, exhibits interference effects but, as we have seen, such an interference is naturally traced out in terms of the normal-mode expansion of the quantized electro-magnetic field (see e.g. the comments in Ref.\cite{xiong_2005}). The result for thermal light in both the signal and the local oscillator ports
shows that the interference pattern is sensitive to the difference in temperature of the two sources.
This is the basic principle used by the FIRAS on board the 
COBE satellite in order to perform high precision measurements of the temperature
and the spectrum of the cosmic micro-wave background radiation.

Since we have been considering initial quantum states in terms of a fixed number of photons as well as ''classical'' states, corresponding to coherent states with an infinite number of photons present, a quantum-mechanical language is mandatory. The signal and local ports of the MMI setup corresponds to independent input sources. It is, of course, a well-known experimental fact that independent photon sources  can give rise to interference effects (see e.g. Refs.\cite{mandel_63,mandel_68, mandel_83,Paul_86,lour_93,Kalten_2006}). Despite the fact that such interference effects are well established, the interpretation of them can, nevertheless,  give rise to  interesting issues regarding the very fundamental aspects of the quantum-mechanical world (see e.g. \cite{glauber_95}) when considering, in particular, interference effects using  single-photon sources. 

We have seen that for multi-mode systems the quantum nature of theses independent sources actually effects the nature of the single-photon intensity measurements. We have already mentioned that the angular-frequency distribution of a single photon can be measured using a similar experimental  setup as  the MMI considered in the present paper \cite{Wasilewski_2007}. With a vacuum state in the local oscillator port and a signal single-photon angular-frequency distribution $f(\omega) $ of the form considered in Eq.(\ref{f_states}), one finds, using Eq.(\ref{fockave}), that
\begin{eqnarray}
\label{one-photon_approx}
\frac{\langle I \rangle (\tau)}{\langle I \rangle (0)} \simeq  \frac{1}{2} \left( 1 +    e^{-(\sigma \tau )^2/4 } \cos\tau\bar{\omega }_s  \right)  \, . ~~~~
\end{eqnarray}
We conclude, in view of our considerations,  by noticing the characteristic exponential behavior in Eq.(\ref{one-photon_approx}) for single-photon interference is not necessarily the indication of a  Lorentzian angular-frequency distribution as claimed in the literature 
when, e.g., measuring the photo-luminenscene signal of a single quantum dot \cite{kammerer_2002} or in studies of other  interference effects of dissimilar photon sources \cite{bennet_2009}.

\noindent \emph{Acknowledgments.} 
This research has made use of NASA's Astrophysics Data System.
The research was supported by the Research Council of 
Norway through grants 170935/V30 and FRINAT-191564,  and by NTNU. One of the authors (B.-S.S.) 
wishes to thank Professor Frederik G.\ Scholtz for a generous and stimulating hospitality 
during a joint  NITheP and Stias, Stellenbosch  (S.A.),  workshop in 2009,  Professors M.\ Reid and D.\ V.\ Ahluwalia, University of Christchurch, N.Z., J.\ R.\ Klauder, University of Florida, Gainesville, U.S.A,  P.\ S.\ Riseborough, Temple University, U.S.A., and the TH-Division at CERN for hospitality when the present work was in progress.

%
%

\end{document}